\documentclass[aps,pre,twocolumn,flosuperscriptaddress]{revtex4-1}
\usepackage{graphicx}
\usepackage{amsmath,amssymb}
\usepackage[utf8]{inputenc}   
\usepackage{color}
\usepackage[colorinlistoftodos]{todonotes}
\usepackage{mathtools, bm}

\newcommand{\be}{\begin{equation}}
\newcommand{\ee}{\end{equation}}
\newcommand{\ba}{\begin{eqnarray}}
\newcommand{\ea}{\end{eqnarray}}

\begin{document}  
\title{Orientational order at finite temperature on undulated surfaces}
\author{Carolina Brito$^1$, Vincenzo Vitelli$^2$ and Olivier Dauchot$^3$,}
\affiliation{$^1$ Instituto de F{\'\i}sica, Universidade Federal do Rio Grande do Sul CP 15051, 91501-970 Porto Alegre RS, Brazil}
\affiliation{$^2$ Institute Lorentz for Theoretical Physics, Leiden University, NL 233 CA, Leiden, The Netherlands}
\affiliation{$^3$ EC2M, UMR Gulliver 7083 CNRS, ESPCI ParisTech, PSL Research University, 10 rue Vauquelin, 75005 Paris, France}
\date{\today}

\begin{abstract}
We study the effect of thermal fluctuations in the XY-model on a surface with non vanishing mean curvature and zero Gaussian curvature. Unlike Gaussian curvature that typically frustrates orientational order, the extrinsic curvature of the surface can act as a local field that promotes long range order at low temperature. We find numerically that the transition from the high temperature isotropic phase to the true long range ordered phase is characterized by critical exponents consistent with those of the flat space Ising model in two dimensions, up to finite size effects. Our results suggest a versatile strategy to achieve geometric control of liquid crystal order by suitable design of the underlying curvature of a substrate or bounding surface. 
\end{abstract}

\maketitle
\section*{Introduction}
The interaction of geometry with orientational order and topological defects determines the spatial organization of many different systems~\cite{NelsonBook}, ranging from the bilayer membranes surrounding cells and intracellular organelles~\cite{NelsonBook2} to the thin-film patterns of block copolymers used to produce nanolithographic masks ~\cite{Park30051997} or supramolecular assembly \cite{Stellacci}.
More broadly, understanding the spatial organization of soft systems under confinement is crucial for the success of new material design strategies based on self-assembly in 2D geometries.

One limitation associated with these systems is the lack of long-range order due to thermal fluctuations: on a flat substrate, two dimensional systems with continuous symmetry and short-ranged interactions don't exhibit long-range order at finite temperature~\cite{MerminWagner}. In principle, this limitation does not hold if the continuous symmetry is broken by coupling the order parameter to one of the principal axis of curvature of a deformed substrate. This geometric interaction acts as an external field that can promote order where none is usually expected. On the other end, the Gaussian (or intrinsic) curvature of a surface leads to geometric frustration that prevents the local order dictated by physical interactions from propagating throughout space~\cite{NelsonBook}.

Theoretical studies of liquid crystals on curved surfaces have primarily focused on determining the ground state texture that emerges from the competition between orientational order and Gaussian curvature~\cite{NelsonBook2, VitelliPRL2004, GiomiAP2009, ParkLubesnky}. However, in experimental systems, the embedding of the substrate in three dimensions is important -- the order parameter (e.g. the nematic director) also couples to the mean, or extrinsic, curvature~\cite{SantangelloPRL2007, VitelliPRE2009, SelingerJPhysChemB2011,VegaSoftMatter2013, Vergori2012}. For instance, in the case of columnar or smectic phases on a curved substrate, the extrinsic curvature can act as an effective field that locally orients the direction of the layers in the ground state ~\cite{SantangelloPRL2007, VitelliPRE2009}. Much less is known about the interplay between ordering fields arising from the extrinsic geometry and thermal fluctuations.

\begin{figure}
\includegraphics[width=3cm, angle=0]{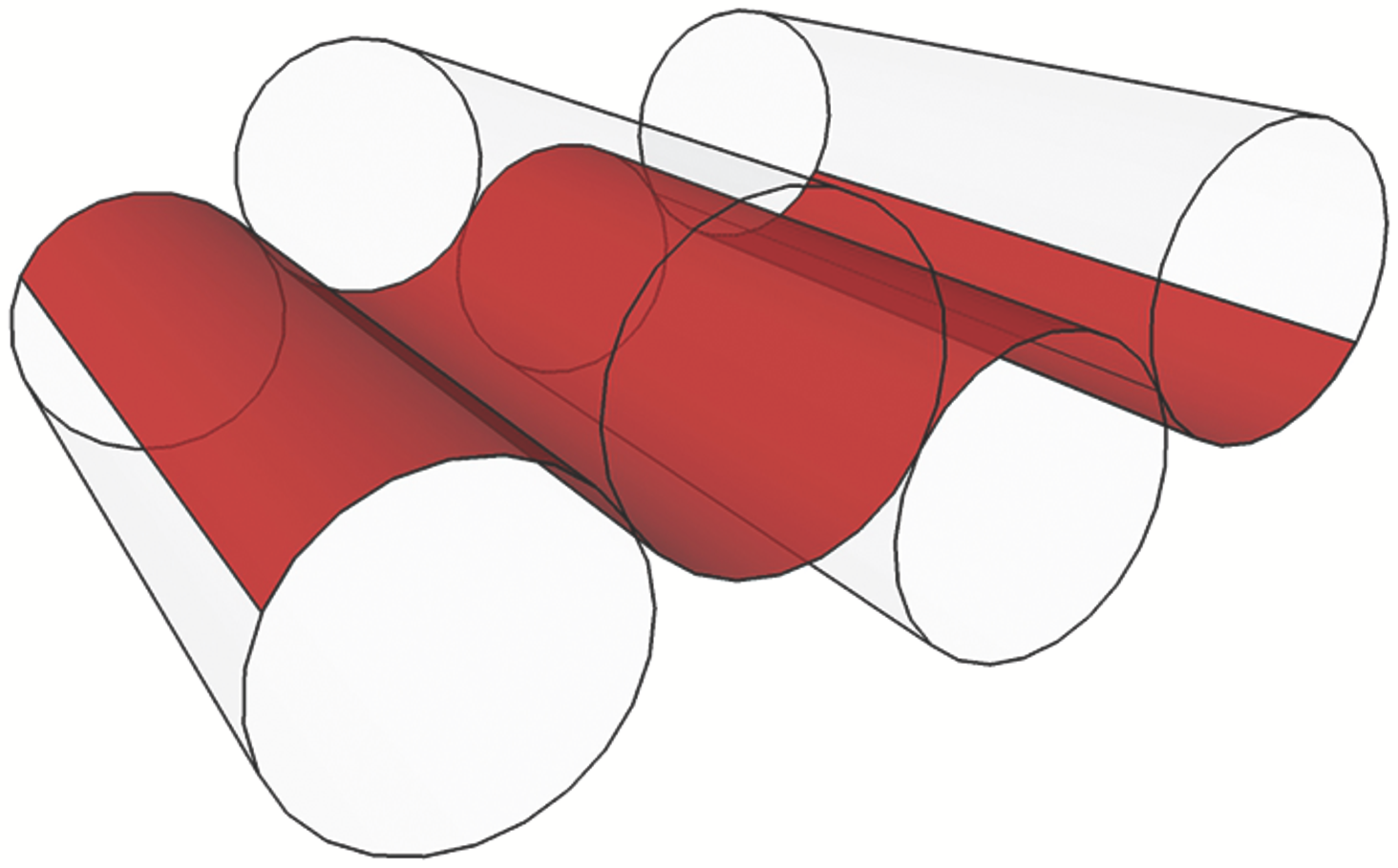}
\includegraphics[trim={0cm 0.5cm 3cm 0.5cm},clip,width=5cm]{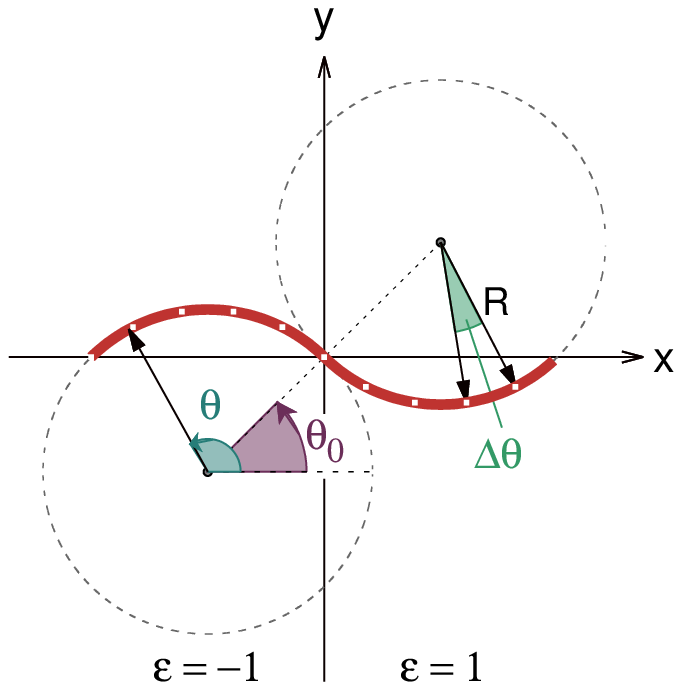}
\caption{{\bf Left:} The XY-model is studied on a curved surface (red) composed of cylindrical sectors alternatively pointing upward and downward.  This surface has zero Gaussian, $\kappa =0$ and constant mean curvature $M$, set by the cylinder radius $R$. {\bf Right:} Cross-section of a unit cell. The cylinders are parallel to the z-direction and tangent at the origin. The red path is the surface where the spins centers -- represented by the white dots -- live.}  
\label{fig:esquemasurface}
\end{figure}

In this paper, we investigate the finite-temperature properties of an isotropic model of in-plane bond orientational order on curved-surfaces with zero Gaussian curvature $\kappa$ but constant mean curvature $M$ (see fig.~\ref{fig:esquemasurface}-left). We propose that orientational order on such curved surfaces can be described as a {\it planar} XY model coupled to an external field which mimics the presence of the extrinsic curvature. This mapping suggests that true long range orientational order can take place at low temperature on a surface of non-vanishing mean curvature. We validate this conclusion by conducting Monte Carlo simulations of the curved XY model and provide quantitative evidence that true long range orientational order is indeed present at low temperature when the extrinsic curvature, that acts as an external field, is non-zero. 
Finite size scaling analysis demonstrates that the critical exponents characterizing the transition from the high temperature isotropic phase to the low temperature ordered phase are consistent, up to numerical accuracy, to those of the planar Ising model.\\

\noindent
\section*{Theoretical background}
 
Consider the Hamiltonian

\begin{eqnarray}
 H_{XY}=-J \sum_{\langle ij \rangle} \vec s_i \cdot \vec s_j
\label{eq:Hxy}
\end{eqnarray}
of the classical XY model that describes unit length spins located on the node of a two dimensional lattice with nearest neighbor interactions of strength $J>0$. This simple model of orientational order exhibits the celebrated Kosterlitz-Thouless transition which is triggered by the unbinding of vortices and anti-vortices \cite{0022-3719-6-7-010, 0022-3719-7-6-005} above the critical temperature $T_{\bm{KT}}$. In the low temperature phase $T<T_{\bm{KT}}$, the vortices are bound, there is only quasi-long range order, and the correlation function of the spins decays algebraically with distance. 
If the spins are linearly coupled to an external field $B_{ \bm{ext}}$, long range order is restored: the transverse spin correlation function decays exponentially at large distance 
$C_{T} \sim e^{-r/\xi_T}$,
with $\xi_T =  (J/B_{ \bm{ext}})^{0.5}$~\cite{deGennes_book}.

In the case of a curved substrate, the in-plane vector order parameter $\bf{s}$(x)$=\cos\theta(x) $$\bf{e_1}$$ + \sin\theta(x) $$\bf{e_2}$ where $\bf{e}_1$ and $\bf{e}_2$ are two orthonormal basis vectors tangent to the substrate and $\theta(x)$ is the local bond angle. Generalization to the case of n-atic order invariant under $\frac{2\pi}{n}$ rotation of the local bond angle is straightforward (e.g. $n=2$ and $n=6$ describe nematic and hexatic order respectively). 
A general continuum elastic energy for n-atic order on a curved substrate reads 
\begin{eqnarray}
H_{\theta}= \frac{K}{2} \int d^2x \sqrt{g}g^{ij} \left(\partial_{i} \theta - A_i \right)\left(\partial_{j} \theta - A_j \right)
\label{eq:H}
\end{eqnarray}
where $K$ is an elastic constant proportional to $n^2$ and the microscopic exchange interaction \cite{ParkLubesnky}. In Eq.~(\ref{eq:H}),  $g_{ij}$ denotes the metric tensor, $g$ is its determinant and $A_i= \bf{e_1}\partial_i \bf{e_2}$ is the spin connection that ensure parallel transport when covariant derivatives of the order parameter are taken. The energy $H_{\theta}$ generalizes the familiar continuum elastic energy of the planar XY model once the substitution $\nabla \theta \rightarrow \partial_{i} \theta - A_i$ is made. It suppresses gradients in the angle $\theta$ that measure the local orientation that the tangent vector makes with respect to its neighbours. At the same time, it captures the geometric frustration induced by the Gaussian curvature of the substrate embedded via the geometric gauge field $A_i$. In writing Eq.~(\ref{eq:H}), anisotropic contributions to $H_{\theta}$ that occurr in the case $n=1$ (classical spins) and $n=2$ (nematic liquid crystals) have been neglected. Note that Eq. \ref{eq:H} needs to be supplemented with a Landau energy functional to account for variations in the {\it amplitude} of the order parameter. 

Recent studies have revealed that the mean curvature of the substrate can play an important role, neglected by $H_{\theta}$, in minimizing gradients of a vector field tangent to a curved surface~\cite{SantangelloPRL2007, VitelliPRE2009, SelingerJPhysChemB2011,VegaSoftMatter2013, Vergori2012}. Upon decomposing the normal curvature of the vector field along the two principal directions of curvature of the surface, an additional contribution to the elastic energy is obtained \cite{SantangelloPRL2007, VitelliPRE2009,SelingerJPhysChemB2011}. 
\begin{eqnarray}
H_{\beta}=  \frac{K}{2} \int d^2x  \sqrt{g} \left[ \kappa_1 ^2 \cos^2 \beta + \kappa_2 ^2 \sin^2 \beta \right]
\label{Kij}
\end{eqnarray}
where $\beta(x)$ represents the angle that the tangent vector forms at position $x$ with respect to the local principal direction with smallest curvature $\kappa_1(x)$ while $\kappa_2(x)$ denotes the largest principal curvature at the same location. 

Eq.~(\ref{Kij}) captures a coupling between bond orientational order and extrinsic curvature whose origin can be intuitively grasped as follows. Consider a cylindrical surface of radius $R$ with a nematic director (or spin) aligned (i) along its azimuthal direction with curvature $\kappa_2=\frac{1}{R}$ or (ii) along its height which has vanishing curvature $\kappa_1 =0$. In both cases, the spins (or molecules) share a common orientation that makes the free energy in Eq.~\ref{eq:H} vanish. However, case (ii) is energetically favorable because in case (i), the molecules are still rotating with respect to each other giving rise to a bending energy proportional to $1/R^2$. Additional source of couplings to the extrinsic curvature, neglected in this study, are possible depending on the symmetries of the order parameter and the specific model under consideration. The continuum elastic theory in Eq.~(\ref{Kij}) suggests that a vector order parameter on a curved surface with zero Gaussian curvature $\kappa=(\kappa_1 \times \kappa_2)=0$ and constant mean curvature $M=\frac{1}{2}(\kappa_1+ \kappa_2)=\frac{1}{2R}$, can be mapped, aside from metric factors, to a flat space XY Hamiltoninan $H_B$ in the presence of an external field: 
\begin{eqnarray}
 H_B=-J \sum_{\langle ij \rangle} \vec s_i \cdot \vec s_j + \sum_{i} B_{i} \sin^2(\beta_i)
\label{Hbext}
\end{eqnarray}
where $\beta_i$ denotes the angle that $\vec s_i$ forms with respect to the external field $B_{i}=J a^2/R^2(\vec{x}_i)$ at position $\vec{x}_i$ and $a$ is the lattice constant of the square grid used to discretize the model. 
A similar mapping holds also for nematic order provided that a coupling of the form $\sin^4(\beta_i)$ is chosen, see Ref. \cite{SantangelloPRL2007, VitelliPRE2009}. Hence, using straightforward trigonometric identies, the finite temperature properties of vector or nematic order on a curved substrate with non vanishing mean curvature are mapped to p-clock models with $p= 2$ or $p=\{2,4\}$ respectively. In both cases one expects to recover the same universality class as the Ising model \cite{PhysRevLett.96.140603, PhysRevB.16.1217b}.\\

\section*{Monte Carlo simulations}

To test this scenario explicitly, we perform Monte Carlo simulations of the XY model described by Eq.~(\ref{eq:Hxy}) on a curved surface of zero Gaussian curvature composed of cylindrical sectors alternatively pointing upward and downward, figure~(\ref{fig:esquemasurface}). It is constructed from the periodical repetition of a unit cell made of two cylinders sectors, parallel to the z-direction. The two cylinders axis are respectively going through the centers $\Omega^{\pm} = (\pm R\cos(\theta_0),\pm R\sin(\theta_0),0)$, with $R$ the cylinders radius, so that the two cylinders are tangent at the origin. The path colored in red in fig.~(\ref{fig:esquemasurface}) is the cross section of our surface of interest.  
In a given cell n, a point of the surface $\vec X$ is parametrized by its coordinates:
\begin{eqnarray}
x&=& x_{O_n} + \epsilon (R\cos(\theta_0) - R\cos(\theta)) \\ \nonumber
y&=& \epsilon (R\sin(\theta_0) - R\sin(\theta)) \\ \nonumber
z&=&z
\end{eqnarray}
where  $x_{O_n}=4(n-1)R\cos(\theta_0)$ is the center of the $\text{n}^{\text{th}}$ cell, where $n = 1~..~\mathcal N_c$.  $\epsilon = 1$, $-1$ when the surface is convex, respectively concave, and $\theta\in[\theta_0,  \pi-\theta_0]$ indicates the angular position of the spin. A spin $\vec s_i$ has coordinates $(s^\theta_i, s^z_i)$ in the tangent frame defined by the unit vectors $\hat e_{\theta}=(\epsilon \sin(\theta ),- \epsilon \cos(\theta))$ and $\hat e_z$. 
The principal curvatures are $\kappa_z = 0$ and $\kappa_\theta = \epsilon/R$. The Gaussian curvature $\kappa = \kappa_z \times \kappa_\theta = 0$, while the mean (or extrinsic) curvature is given by $M = (\kappa_z+\kappa_\theta)/2= \epsilon/2R$. 
In each unit cell the number of spins is $N_{\theta} \times N_z$. Spins are regularly spaced on a square lattice, of mesh size $a = R \Delta\theta$, with $\Delta\theta = 2(\pi - 2\theta_0)/N_\theta$. Choosing $a$ as the unit length scale, the curvature $M=a/2R=\Delta\theta/2$  is continuously tuned by changing $R$. $\theta_0$ follows in order to keep $N_\theta$ constant. In the following $N_z = N_\theta$, so that the total number of spins per unit cell is always $ N_\theta^2$.
Finite size effects are studied by increasing the number of unit cells in both $x$ and $z$ directions, leading to  systems of linear size $L= \mathcal N \times N_\theta \times a$ and total number of spins $N = (\mathcal N N_\theta)^2$, where $\mathcal N$ is the linear number of unit cells. 
The above setting has the advantage to allow for a separate analysis of the effects of curvature and system size. 
\begin{figure}
\includegraphics[width=8.7cm]{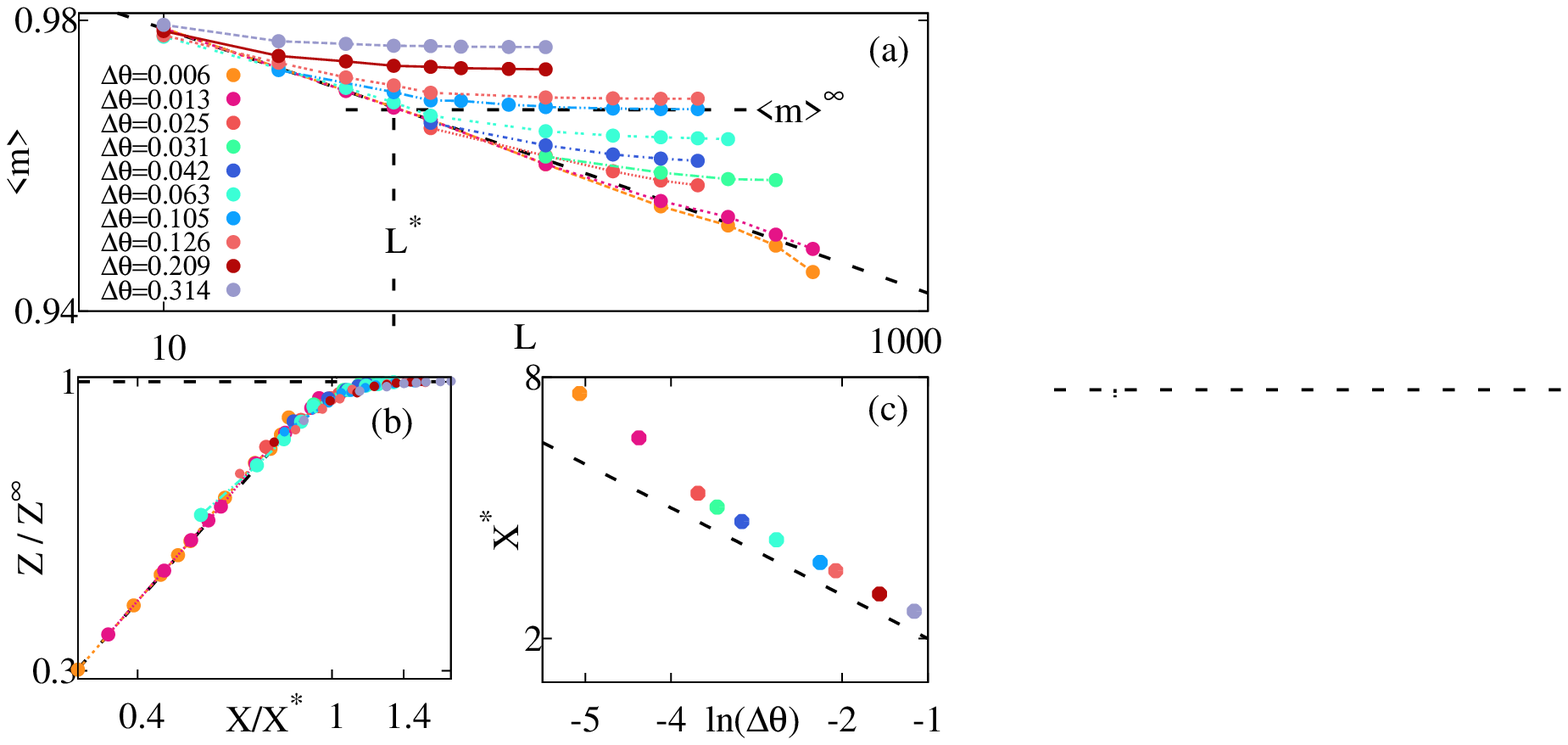}
\caption{{{\bf Magnetization at low T:} {\bf (a)} $\langle m\rangle$ as function of the system size $L$ for various curvature $\Delta\theta$.  Each point is an average over at least 100 initial conditions. The dashed line is the theoretical prediction for the zero curvature case  (eq.~\ref{eq:flat}) {\bf (b)} $Z/Z^{\infty}$ {\it vs} $X/X^{*}$. The dashed line is  $Z/Z^{\infty} = X/X^{*}$.   {\bf (c)}  $X^{*}$  {\it vs} $\ln(\Delta \theta)$. The dashed line indicates slope $-1$.}}
\label{scaling}
\vspace{-3mm}
\end{figure}

After initializing the spin orientations according to a flat distribution, we use Monte Carlo simulations to minimize the energy of the system, following the single-spin-flip  Metropolis algorithm \cite{bookNewman}.
We measure the averaged magnetization $\left<m\right>=\left<\left||\sum_i \vec s_i\right||\right>$, with the expression for $m$ being:
\be 
m =  \frac{\sqrt{(\sum_i  s_i^z)^2+(\sum_i  s_i^{\theta}~ sin(\theta_i))^2 + (\sum_i  s_i^{\theta}~ cos(\theta_i))^2}}{N},  \nonumber 
\ee
its fluctuations and the transverse correlation $C_T(r)$:
\be
C_T(r)= \langle  s_i^{\theta}(\vec r)\hat e_{\theta}^i \cdot  s_j^{\theta} (\vec r+\vec r')\hat e_{\theta}^j \rangle,  \nonumber 
\nonumber 
\ee
where the last average $\langle \cdot \rangle$ runs over the spins that are distant by $r$. 
All measurements are made after the system reaches its asymptotic equilibrium state (usually $10^5$ iterations per spin) and  averaged over at least 100 initial conditions. 
We consider several system sizes with $N_\theta = 10, 20, 50, 100 $ and $\mathcal N=1, 2, 3, 4, 5$, that is a number of spins spanning from $N=(1 \times 10)^2$ up to $N=(5 \times 100)^2$.\\ 
%

\section*{Results}

We first consider the low temperature regime, $T=0.1<<T_{\rm{KT}}$. For finite but not too large system size, the XY-model in a plane has a finite magnetization which decreases logarithmically with system size:
\begin{eqnarray}
\left<m\right>(T,L) = \left<m\right>_0(T) - \frac{k_B T}{4\pi J} \ln(L/a),
\label{eq:flat}
\end{eqnarray}
with $\left<m\right>_0(T)\simeq 1$ when  $T=0.1<<T_{\rm{KT}}$ .  Figure~\ref{scaling}(a) shows how the introduction of the extrinsic curvature qualitatively modifies this behavior. When $\Delta\theta>0$, the magnetization decreases with system sizes following eq.~(\ref{eq:flat}) until, for $L>L^*$, it saturates to a constant value $\left<m\right>^{\infty}$, to which it tends to, when $L/a \rightarrow \infty$. It is easy to verify that the magnetization obeys a scaling relation $\left<m\right> (\Delta\theta, L) = \tilde m (L/L^*(\Delta\theta))$, which is best illustrated in figure~\ref{scaling}(b) introducing $Z=\left<m\right>_0 - \left<m\right> $ and $X=\ln(L/a)$, $Z^{\infty} = \left<m\right>_0 - \left<m\right>^{\infty}$ and $X^* = \ln(L^*/a)$. From this scaling, one extracts the dependence of $L^*/a$ as a function of $\Delta\theta$ plotted in figure~\ref{scaling}(c). For large enough $\Delta\theta$, $L^*/a \simeq \Delta\theta^{-1} = R/a$ : the crossover length scale is simply the curvature radius. One also obtains the \emph{finite} value of the magnetization in the thermodynamic limit, $\left<m\right>^{\infty} = \left<m\right>_0 - \ln(L^*(\Delta\theta)/a) \simeq {\rm Cst} +  b \ln(\Delta\theta))$. 
\begin{figure}
 \includegraphics[width=8.7cm]{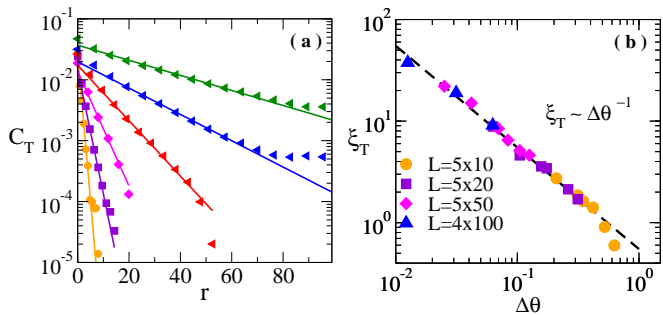}
\caption{{\bf Transversal correlations:} {\bf (a)} $C_T(r)$ for different $\Delta\theta = [0.52, 0.26, 0.125, 0.063, 0.031, 0.013]$ and different system sizes (see legend in {\bf (b)}). {\bf (b)}  Transversal correlation length $\xi_T$, extracted from exponential fits as shown in {\bf (a)}, as a function of $\Delta \theta$. The dashed line is a power-law fit with: $\xi_T \sim \Delta \theta^{-1}$.}
\label{Correl_xiT_models_I_II}
\vspace{-3mm}
\end{figure}
\begin{figure*}[t!]
\includegraphics[width=8.5cm]{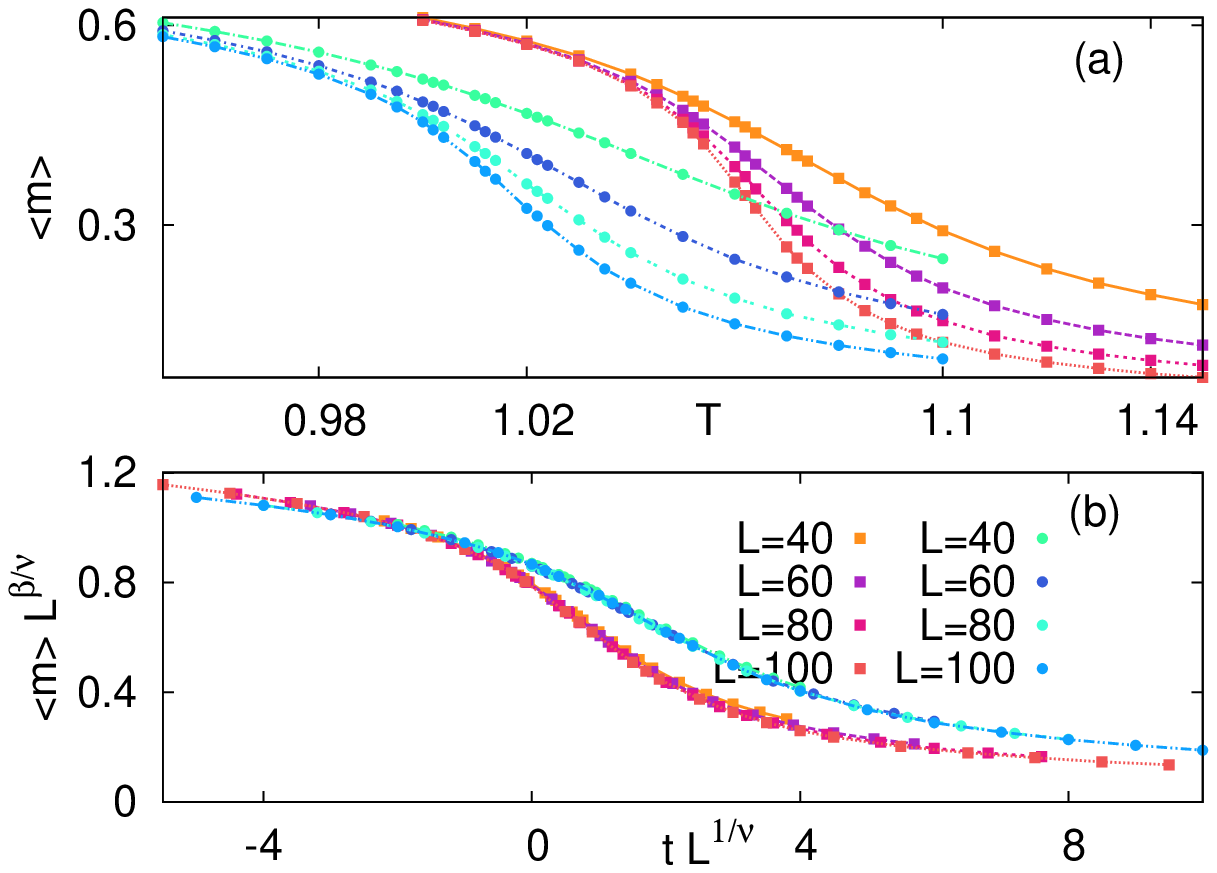}
\includegraphics[width=8.5cm]{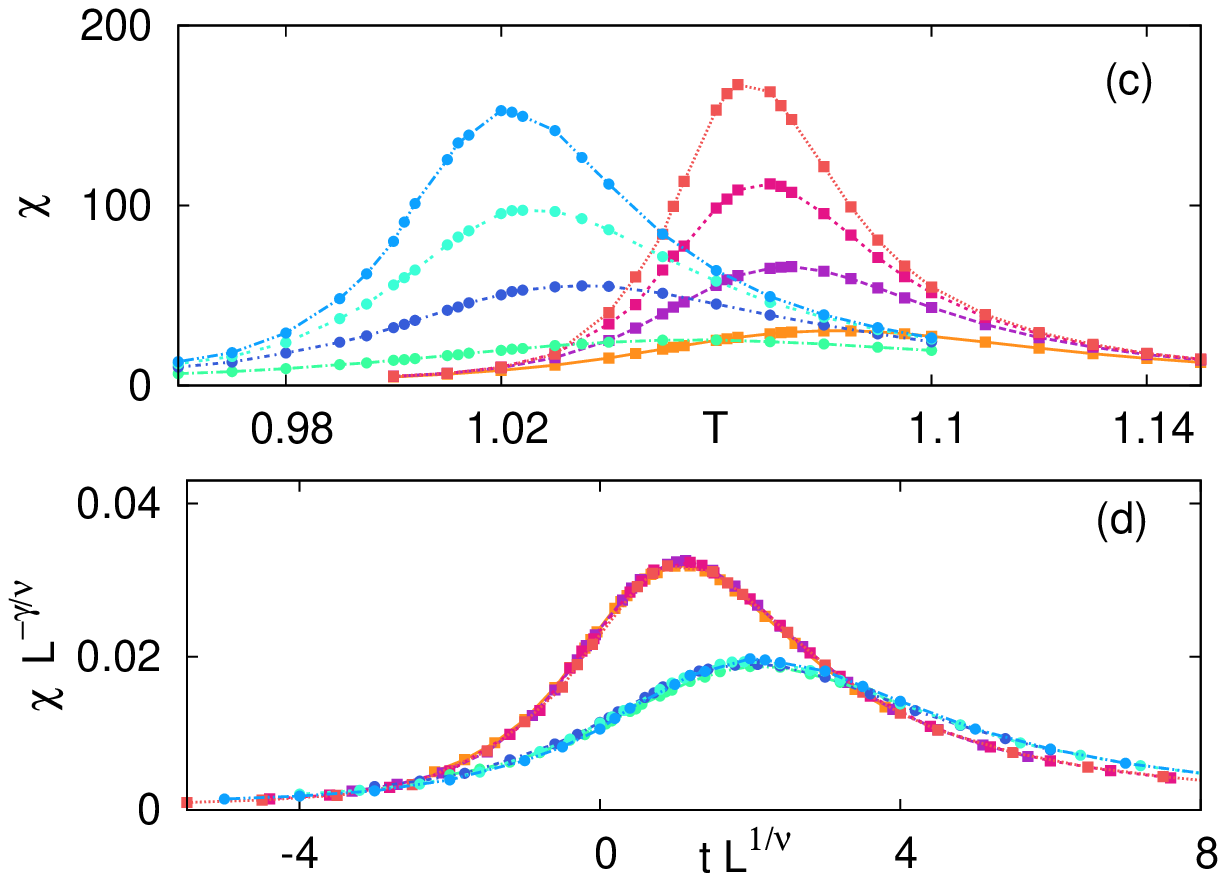}
\caption{ {\bf Magnetization $\langle m \rangle$ and magnetic susceptibility $\chi$ as a function of temperature.}
{\bf (a)} $\langle m \rangle$ {\it vs} $T$ (zoom on the transition). {\bf (b)} same data as {\bf (a)} with $\langle m \rangle$ rescaled by $L^{\beta/\nu}$ and $t=(T-T_c)/T_c$ rescaled by $L^{1/\nu}$. {\bf (c)} $\chi$ {\it vs} $T$ (zoom on the transition).  {\bf (d)} same data as {\bf (c)} with $\chi$ rescaled by $L^{-\gamma/\nu}$ and $t=(T-T_c)/T_c$ rescaled by $L^{1/\nu}$. Squares are numerical data on the curved surface with fixed curvature $\Delta\theta_1=\pi/10$ and circles correspond to a curvature of $\Delta\theta_2=\pi/20$. The rescaling exponents are summarize in table I.}  
\label{curvBext_all}
\end{figure*}
Finally, the transverse correlations $C_T(r)$ of the magnetization fluctuations decreases exponentially (fig.~\ref{Correl_xiT_models_I_II}-(a)) and the transverse correlation length $\xi_T$  scales like the inverse mean curvature
\be
\xi_T\sim \Delta \theta ^{-1} \sim R.
\ee
Recalling that for the planar XY model in the presence of an external field, one has $~\xi_T \sim B_{ext} ^{-0.5}$~\cite{deGennes_book}, the above results confirm the mapping anticipated theoretically between the extrinsic curvature and an external field with amplitude inversely proportional to $R^2$. 
We have thus shown that the XY model on a curved surface with zero Gaussian and finite mean curvatures exhibits a phase with true long range magnetic order at low temperature.

We now analyze what kind of transition happens at fixed mean curvature when the temperature is increased. Figure (\ref{curvBext_all}) shows the average magnetization $\langle m \rangle$  and the  magnetic susceptibility $\chi = \beta N_T (\langle m^2 \rangle-\langle m \rangle^2)$ as a function of temperature $T$ for different system sizes as well as their finite size rescaling across the transition. We observe the following phenomenology: (i) when temperature increases, the magnetization decreases and goes to zero above a given temperature, (ii) the magnetic susceptibility shows a sharp increase around the temperature where $\langle m \rangle$ goes to zero and (iii) when the system sizes increase, $\chi$ becomes sharper. 
Since the curvature (akin to an external field) breaks the continuous symmetry of the spins, the system has now only two favored states. We then expect a phase transition similar to the Ising model \cite{bookYeomans}. To investigate this transition, we define the reduced temperature $t = (T-T_c)/T_c$  and investigate the following finite size scalings \cite{PhysRevE.89.012126}:
\begin{eqnarray}
\langle m \rangle =& L^{-\beta/ \nu}  \tilde m ( L^{-1/\nu}t),\\
\label{eq_m_resc}
\chi =& L^{\gamma/ \nu}  \tilde \chi ( L^{-1/\nu}t),\\
\label{eq_xi_resc}
C =& L^{\alpha/ \nu}  \tilde C ( L^{-1/\nu}t),
\label{eq_c_resc}
\end{eqnarray}
where $C=\beta^2 N_T (\langle e^2 \rangle-\langle e \rangle^2)$ is the specific heat with $\langle e \rangle$ the average energy per spin.
We find $T_c=1.055$, and $T_c=1.000$ for the curvature $\Delta\theta_1=\pi/10$, and $\Delta\theta_2=\pi/20$ respectively. 
The study of the heat capacity, not shown here, reveals very small values for $\alpha$, suggesting the existence of a logarithmic singularity with $\alpha = 0$, hence $\nu =1$, as it is the case in the 2d Ising model. In the following, we take those values for granted and perform the finite size scaling analysis on the magnetization and the susceptibility to obtain the exponents $\beta$ and $\gamma$, which are summarized in the table \ref{tab_exp} together with the exponents of the Ising model in two dimensions. 
\begin{table}
\begin{center}
\begin{tabular}{ || c || c | c | c ||} \hline \hline
exponent &  $\Delta\theta_1$ & $\Delta\theta_2$ &  Ising  \\ \hline \hline
  $\beta$   & [0.11 0.17]  & [0.11 0.17]   & 0.125   \\ \hline
  $\gamma$  & [1.80 1.90]  & [1.95 2.05]   & 1.75    \\ \hline
\end{tabular}
\caption{Estimated values of the critical exponents for two curvature $\Delta\theta_1 = \pi/10$, $\Delta\theta_2 = \pi/20$ and exact values for the Ising model in two dimensions.}
\label{tab_exp}
\end{center}
\end{table}
Within numerical accuracy, we find consistent values for the critical exponent $\beta$ for the two curved surface systems and those are compatible with the 2d Ising value for $\beta$. On the contrary, the exponent $\gamma$ depends slightly on the curvature: the smaller the curvature, the larger $\gamma$. Also the 2d Ising model value for $\gamma$ does not match our numerical data in the range of system sizes explored here. In practice $L$ is never very much larger than the crossover size $L^*$, above which the system starts probing the curvature (see fig.(\ref{scaling})). We thus attribute our observations to the very strong finite size effects, associated with the essential singularity of the correlation length when approaching the KT transition from the high temperature phase~\cite{PhysRevB.55.3580}. 
Also, as the curvature is increased, the effect of the metric factors in Eqs. (\ref{eq:H}) and (\ref{Kij}), neglected in our mapping, becomes more important. 
This difference notwithstanding, our results suggest that in the thermodynamic limit, the XY-model on the curved surface belongs to the 2d-Ising universality class. 

\section*{Discussion and Conclusions}
In this work, we studied the interaction between the extrinsic geometry of a substrate and the orientational order of the system in the presence of thermal fluctuations. Using theoretical arguments and Monte Carlo simulations of a classical XY-model confined to a surface with zero Gaussian and finite mean curvature, we showed  that: (i) the XY-model on the curved space can be mapped to the XY-model on a flat space coupled to an external field, (ii) the mean curvature acts as an external field and generates a true long order phase at  low temperature and (iii) when temperature increases, the system experiences a critical phase transition akin to the 2D Ising model. 

Our results represent a special case of a general phenomenon that is also behind the famous stabilization of long-range crystalline order by out of plane thermal fluctuations in two dimensional solid membranes \cite{NelsonBook2}. Here, we show that the stabilization of long-range (orientational) order persists even when the underlying surface is frozen and restricted to have only zero mean curvature. In this case, the qualitative effect of a slowly varying radius of curvature $R(x)$ on thermal fluctuations can be captured by mapping it to an external field whose local magnitude is inversely proportional to $R^2(x)$. This simple correspondence provides an intuitive design criterion to experimentally realize  surface morphologies that achieve spatial control or modulation of in-plane orientational order in liquid crystal monolayers.  

\section*{Acknowledgements}
We thank the authors of\cite{PhysRevE.89.012126} for very useful discussions about the FSS analysis and G. Canova for sharing his data of $XY$-model on the plane with us.  We also thank G. Tarjus, R. Kamien and A. Souslov for helpful comments and discussions.

\bibliographystyle{apsrev4-1}
\bibliography{xy_references}

\end{document}